\documentclass[
  aps,
  prl,
  reprint,
  preprintnumbers,
  showpacs,
  groupedaddress,
  amsmath,
  amssymb,
  floatfix]{revtex4-1}
\usepackage{graphicx,epsfig,dcolumn,multirow}
\usepackage{natbib}
\newcolumntype{d}[1]{D{.}{.}{#1}}

\RequirePackage{xspace}

\newcommand{\simgt}{\,\hbox{\lower0.6ex\hbox{$\sim$}\llap{\raise0.6ex\hbox{$>$}}}\,}
\newcommand{\simlt}{\,\hbox{\lower0.6ex\hbox{$\sim$}\llap{\raise0.6ex\hbox{$<$}}}\,} 

\newcommand{\ov}[1]{\bar{#1}}

\newcommand{\fig}[1]{Fig.~\ref{#1}}

\newcommand{\gev}{\ensuremath{\,\text{GeV}}}

\newcommand{\Bcal}{\mathcal{B}}

\usepackage{pstricks}
\newcmykcolor{darkgreen}{1 0 0.6 0.5}  %% cyan magenta yellow black

\newcommand{\myFitter}{\textit{my}Fitter}

\begin{document}
%% \linenumbers

\preprint{TTP12-023, CERN-PH-TH-2012-182}

\title{Status of the fourth fermion generation before ICHEP2012:\\
  Higgs data and electroweak precision observables}

\author{Otto Eberhardt$^{\,a}$, 
   % Geoffrey Herbert$^{\,b}$, 
   %% Heiko Lacker$^{\,b}$,\\ 
  Alexander Lenz$^{\,b}$,  Andreas Menzel$^{\,c}$,\\ Ulrich 
  Nierste$^{\,a}$, and Martin Wiebusch$^{\,a}$  
\vspace{0.6cm}}

\affiliation{
\mbox{$^{a}$ Institut f\"ur Theoretische Teilchenphysik,
Karlsruhe Institute of Technology, D-76128 Karlsruhe, Germany,}
\mbox{email: otto.eberhardt@kit.edu, ulrich.nierste@kit.edu, 
              martin.wiebusch@kit.edu}\\                
\mbox{$^{b}$ CERN - Theory Divison, PH-TH, Case C01600, CH-1211 Geneva 23,
                {e-mail: alenz@cern.ch}}\\
\mbox{$^{c}$  Humboldt-Universit\"at zu Berlin,
                   Institut f\"ur Physik,
                   Newtonstr. 15,
                   D-12489 Berlin, Germany,}
\mbox{e-mail: % geoffrey.herbert@physik.hu-berlin.de, 
              % lacker@physik.hu-berlin.de,
              amenzel@physik.hu-berlin.de}\\
}  
\date{\today}

\begin{abstract}
  We perform a global fit of the parameters of the Standard Model with a
  sequential fourth generation (SM4) to LHC and Tevatron Higgs data and
  electroweak precision data.  Using several likelihood ratio tests we compare
  the performance of the SM4 and SM3 at describing the measured data.  Since the
  SM3 and SM4 are not \emph{nested} (i.e.\ the SM3 can not be considered as a
  special case of the SM4 with some parameters fixed) the usual analytical
  formulae for $p$-values in likelihood ratio tests do not hold. We thus apply a
  new method to compute these $p$-values. For a Higgs mass of $126.5\gev$ and
  fourth-generation quark masses above {600\gev} we find that the SM4 is
  excluded at $3.1\sigma$.
\end{abstract}

%========================================

\maketitle
\section{Method and inputs}
In this paper we study the SM4, which differs from the established Standard
Model (denoted by SM3) by an additional fermion generation. We treat the masses
of the extra fermions as free parameters and allow for arbitrary flavor mixings
among the quarks of the four generations in our fits. Large mixings of the
fourth-generation lepton doublet with those of the first three generations are
ruled out \cite{Lacker:2010zz} from data on lepton-flavor violating decays and
lepton-flavor universality \cite{NA62:2011aa}. Recent NA62 data constrain these
mixing angles even further \cite{NA48/2:2012gh}. Including lepton mixing within
the allowed range has a negligible impact on the electroweak precision
observables (EWPOs). In the absence of lepton mixing the decay of the Higgs
boson into neutrinos is invisible as long as the fourth-generation charged
lepton is heavier than the corresponding neutrino. This invisible Higgs decay
mode increases the total Higgs width and potentially counterbalances the effect
of the enhanced $gg\to H$ production mechanism \cite{Gunion:1995tp,
  Kribs:2007nz}, because the branching fractions into the observed final states
are reduced \cite{Cetin:2011fp,Eberhardt:2012sb}. Allowing for (even small)
mixing of the fourth with the other lepton doublets can render the neutrino
decay mode visible. Since we want to quantify the level at which the SM4 is
ruled out, we may confine ourselves to the most conservative scenario with an
unmixed fourth-generation lepton doublet. Like the SM3, the SM4 can be studied
with Dirac or Majorana neutrinos. In the fits presented in this paper we use
Dirac neutrinos. In our conclusions we briefly discuss the (marginal) changes in
the results expected for Majorana neutrinos. From a model-building point of
view, the hierarchy between three almost massless neutrinos and a fourth
neutrino with mass of order of the electroweak scale can be motivated by a
symmetry enforcing massless neutrinos in the exact symmetry limit: e.g.\ three
right-handed neutrino fields might carry some U(1) charge while the fourth
neutrino field and the left-handed lepton doublets are uncharged under this new
symmetry. The Yukawa couplings are small spurions breaking this symmetry,
leading to three tiny neutrino masses and tiny mixings between the fourth and
the other generations.

A sequential fourth generation of fermions decouples neither from the production
cross section $\sigma(gg \to H)$ nor from the Higgs decay rate into
photons. Consequently, current LHC Higgs data put the SM4 under serious pressure
\cite{Djouadi:2012ae, Kuflik:2012ai, Eberhardt:2012sb, Buchkremer:2012yy}. In a
recent publication \cite{Eberhardt:2012sb} we presented a global fit of the SM4
parameters to EWPOs and Higgs signal strengths measured at Tevatron and the
LHC. The signal strength is defined as
\begin{equation}\label{eq:signalstrength}
  \hat\mu(X\to H\to Y) = \frac{\sigma(X\to H)\Bcal(H\to Y)|_\text{SM4}}
                              {\sigma(X\to H)\Bcal(H\to Y)|_\text{SM3}}  
  \quad.
\end{equation}
Here we update our results with all available data and analyse the status of the
SM4 prior to the ICHEP2012 conference. We also compute the statistical
significance ($p$-value) at which the SM4 is excluded. As explained in
\cite{Eberhardt:2012sb} the computation of the $p$-value is non-trivial: due to
the non-decoupling nature of the fourth-generation fermions the SM3 can not be
regarded as a special case of the SM4, i.e. the two models are not
\emph{nested}. Analytical formulae for $p$-values only hold for nested models
and thus the $p$-value of the SM4 has to be computed numerically. To this end, a
new C++ framework for maximum likelihood fits and likelihood ratio tests called
{\myFitter} \cite{myFitter-page} was written. The implementation is discussed in
\cite{myFitter}.

In total, the following aspects of our previous analysis have been improved:
\begin{enumerate}
\item The masses of all four fourth-generation fermions are now consistently
  treated as free parameters.  To avoid non-perturbative Yukawa couplings and
  constraints from direct searches of fourth-generation quarks we require
  $600\gev\leq m_{t'},m_{b'}\leq 800\gev$. We are aware that
    for fermion masses of $800\gev$ the validity of perturbation theory is
    questionable at best. However, reducing the upper limit for the fermion
    masses can only lead to larger $\chi^2$ values in the SM4 and thus to
    smaller $p$-values. In this sense, the upper limit of $800\gev$ is a
    \emph{conservative} estimate.
\item The signal strength for $pp\to H\to\tau\tau$ measured at the LHC
  \cite{ATLAS-CONF-2012-019} is included in the analysis.
\item In the global fit, the Higgs mass is no longer fixed at $125\gev$, but is
  allowed to float in the range where experimental data on the Higgs signal
  strengths is available, i.e.\ $115\gev\leq m_H\leq 150\gev$.\footnote{A
    lattice study has found the lower bound $m_H \gtrsim 500\gev$ for
    $m_{t'}=m_{b'}=700\gev$ \cite{Gerhold:2010wv}. We interpret this result such
    that the perturbative vacuum state is metastable for $m_H \approx 125 \gev$
    and the heavy quark masses used by us. Therefore Ref.~\cite{Gerhold:2010wv}
    per se does not invalidate our analysis.}
\item Since, for a variable Higgs mass, no separate $H\to\gamma\gamma$ signal
  strengths for the gluon fusion and vector boson fusion production modes are
  available we only use the combined signal strength for $pp\to
  H\to\gamma\gamma$ as input.
\item For the two cases $m_H=126.5\gev$ (the preferred Higgs mass of
  the SM3) and $m_H=147\gev$ (the preferred Higgs mass of the SM4) we
  perform likelihood ratio tests to compare the performance of the SM3 and SM4
  at describing the measured data.
\end{enumerate}
Regarding the last point, a few more comments are in order. In likelihood ratio
tests the difference $\Delta\chi^2$ of minimal $\chi^2$ values obtained in the
SM3 and the SM4 is used as a test statistic. One then assumes that the measured
observables are random variables distributed around the prediction of one model
(e.g.\ the SM4) with a spread determined by their errors and computes the
probability ($p$-value) that a random set of ``toy-observables'' leads to a
$\Delta\chi^2$ which is more extreme (e.g.\ more SM3-like) than the
$\Delta\chi^2$-value obtained from the real data. Note that this is different
from the goodness-of-fit analysis presented in \cite{Kuflik:2012ai}, which used
the $\chi^2$ value of the SM4 as a test statistic and therefore did not compare
the performance of the SM3 and the SM4. Also, the $H\to\tau\tau$ signal
strengths were not included in their analysis.

Unfortunately, the likelihood ratio tests can not be done (by us) if the Higgs
mass is treated as a free parameter.  In that case, the signal strengths
measured in each invariant mass bin of each Higgs decay mode would have to be
treated as separate observables, and we do not have any information on
  statistical correlations between adjacent bins. Thus we only perform
likelihood ratio tests for \emph{specialisations} of the SM3 and SM4, where the
Higgs mass is fixed to $m_H=126.5\gev$ (the value preferred by the global SM3
fit) or $m_H=147\gev$ (the value preferred by the global SM4 fit). Then only the
signal strengths at $m_H=126.5\gev$ and $m_H=147\gev$ have to be treated as
independent observables and correlations between these observables can safely be
neglected.

Note, however, that the information from \emph{all} invariant mass bins is
encoded in our $\chi^2$ function. So, for example, the $\chi^2$ value at
$m_H=147\gev$ has a contribution due to the fact that there is a signal at
$m_H=126.5\gev$.  If the model under consideration had a Higgs boson
outside the discovery reach of LHC (or no Higgs boson at all), the theory
prediction for all signal strengths in all invariant mass bins would be
zero. This leads to a constant contribution to the $\chi^2$, which we are
allowed to drop.  Now assume that the model has a Higgs boson with some mass
$m_H$ and a predicted signal strength $\hat\mu_\text{th}(m_H)$. Let
$\hat\mu_\text{ex}(m_H)$ and $\Delta\hat\mu(m_H)$ be the measured signal
strength and experimental error for the corresponding invariant mass bin. After
dropping the constant, the $\chi^2$ function is
\begin{equation}
  \chi^2(m_H)=
  \frac{[\hat\mu_\text{th}(m_H)-\hat\mu_\text{ex}(m_H)]^2
        - [\hat\mu_\text{ex}(m_H)]^2}{[\Delta\hat\mu(m_H)]^2}
  \quad.
\end{equation}
If there is a clear signal at the Higgs mass $m_H$, the second term gives a
large negative contribution to the $\chi^2$ function. This contribution is
not present if $m_H$ is in a region without a signal, so the minimum of the
$\chi^2$ function will usually be at a Higgs mass close to the signal.

In the present analysis, the following experimental inputs are used:
\begin{itemize}
\item[i)] $\hat\mu(pp\to H\to WW^*)$ measured by ATLAS \cite{ATLAS:2012sc},
\item[ii)] $\hat\mu(pp\to H\to\gamma\gamma)$ 
  measured by ATLAS \cite{ATLAS:2012ad},
\item[iii)] $\hat\mu(p\bar p\to HV\to Vb\ov b)$ measured by CDF and D0 
     \cite{Knoepfel:2012xk},
\item[iv)] $\hat\mu(pp\to H\to ZZ^*)$ and 
  $\hat\mu(pp\to H\to \tau \ov{\tau})$ measured by ATLAS
  \cite{ATLAS-CONF-2012-019},
\item[iv)] the electroweak precision observables (EWPOs) $M_Z$, $\Gamma_Z$,
  $\sigma_\text{had}$, $A_\text{FB}^l$, $A_\text{FB}^c$, $A_\text{FB}^b$, $A_l$,
  $A_c$, $A_b$, $R_l=\Gamma_{l^+l^-}/\Gamma_\text{had}$, $R_c$, $R_b$,
  $\sin^2\theta_l^\text{eff}$ measured at LEP and SLC \cite{EWWG:2010vi} as well
  as $m_t$, $M_W$, $\Gamma_W$ and $\Delta\alpha_\text{had}^{(5)}$
  \cite{Nakamura:2010zzi}.
\item[v)] the lower bounds $m_{t^\prime,b^\prime}\gtrsim 600\;$\gev
  (from the LHC) \cite{Aad:2012xc, Aad:2012us, CMS:2012ye, CMS-PAS-EXO-11-099}
  and $m_{l_4}>101\,\text{GeV}$ (from LEP2) \cite{Nakamura:2010zzi}.
\end{itemize}
Unfortunately, there is no data for signal strengths as a function of the Higgs
mass from CMS.

On the theory side, the global fits with a variable Higgs mass were done with
the CKMfitter software \cite{Hocker:2001xe}. The EWPOs in the SM4 were
calculated with the method described in \cite{Gonzalez:2011he}, using FeynArts,
FormCalc and LoopTools \cite{Hahn:1998yk, Hahn:2000kx, Hahn:2006qw} to compute
the SM4 corrections to the EWPOs. The EWPOs in the SM3 were calculated with the
ZFitter software \cite{Bardin:1989tq, Bardin:1999yd, Arbuzov:2005ma}. The Higgs
width and branching ratios in the SM4 and SM3 were calculated with HDECAY
v.\ 4.45 \cite{Djouadi:1997yw}, which implements results of
\cite{Djouadi:1994gf, Djouadi:1994ge, Passarino:2011kv, Denner:2011vt}. The SM3
Higgs production cross sections were taken from \cite{Dittmaier:2011ti} (LHC)
and \cite{Brein:2003wg, Baglio:2010um} (Tevatron). For the numerical integration
required to compute the $p$-values we use the Dvegas code \cite{Dvegas} which
was developed in the context of \cite{Kauer:2001sp, Kauer:2002sn}.

%%%%%%%%%%%%%%%%%%%%%%%%%%%%%%%%%%%%%%%%%%%%%%%%%%%%%%%%%%%%%%%%%%%%%%%%%%%%%%% 
\section{Results}
%%%%%%%%%%%%%%%%%%%%%%%%%%%%%%%%%%%%%%%%%%%%%%%%%%%%%%%%%%%%%%%%%%%%%%%%%%%%%%% 

\begin{figure}[t]
  \includegraphics[width=0.45\textwidth,bb=150 480 486 709,clip=true]{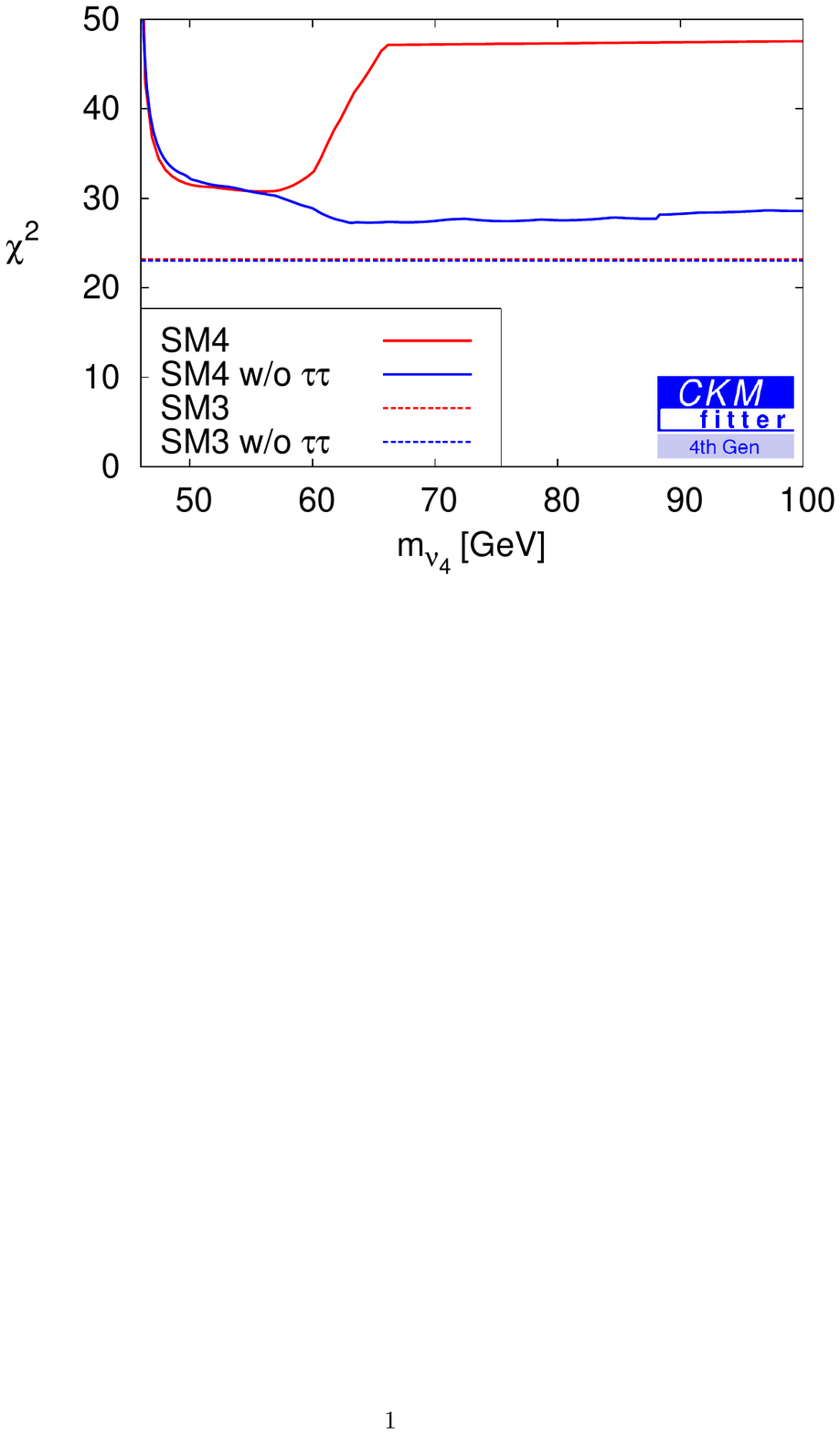}
  \caption{Minimum $\chi^2$ values for a fixed neutrino mass as a function of
    $m_{\nu_4}$. The blue (red) lines show the results from the combined
    analysis of EWPOs and Higgs signal strengths with (without) the
    $H\to\tau\tau$ channel. The solid and dashed lines correspond to the
    SM4 and SM3, respectively.}
  \label{fig:mnu}
\end{figure}

To show the impact of the $H\to\tau\tau$ signal strength we plot the minimal
$\chi^2$ value with and without the $H\to\tau\tau$ input as a function of the
mass $m_{\nu_4}$ of the fourth-generation neutrino in Fig.~\ref{fig:mnu}. We see
that for $m_{\nu_4}\lesssim 60\gev$ the minimum $\chi^2$ values are almost the
same with and without the $H\to\tau\tau$ input. For $m_{\nu_4}\gtrsim 65\gev$
the $H\to\tau\tau$ input increases the minimum $\chi^2$ by more than 20. We also
see that without the $H\to\tau\tau$ input the SM4 favours large values of
$m_{\nu_4}$.With the $H\to\tau\tau$ signal strengths included, the smallest
$\chi^2$ values are obtained for $m_{\nu_4}$ between $50$ and $60\gev$.

This can be understood as follows: the production rate of Higgs bosons in gluon
fusion is enhanced by a factor of 9 in the SM4 due to the contributions from
additional heavy quark loops. On the other hand, the effective $HWW$, $HZZ$ and
$H\gamma\gamma$ couplings are suppressed by the higher order corrections
discussed in \cite{Denner:2011vt}. No such suppression is possible for
$H\to\tau\tau$, so we would expect a $H\to\tau\tau$ signal strength of $9$.  The
only way to reduce this signal strength is to open the invisible
$H\to\nu_4\bar\nu_4$ decay mode, which then suppresses all branching ratios by a
common factor. Thus, for large values of $m_{\nu_4}$, the fit gets considerably
worse if the $H\to\tau\tau$ channel is included.

\begin{figure}
  \includegraphics[width=0.45\textwidth,viewport=150 480 486 705,clip]{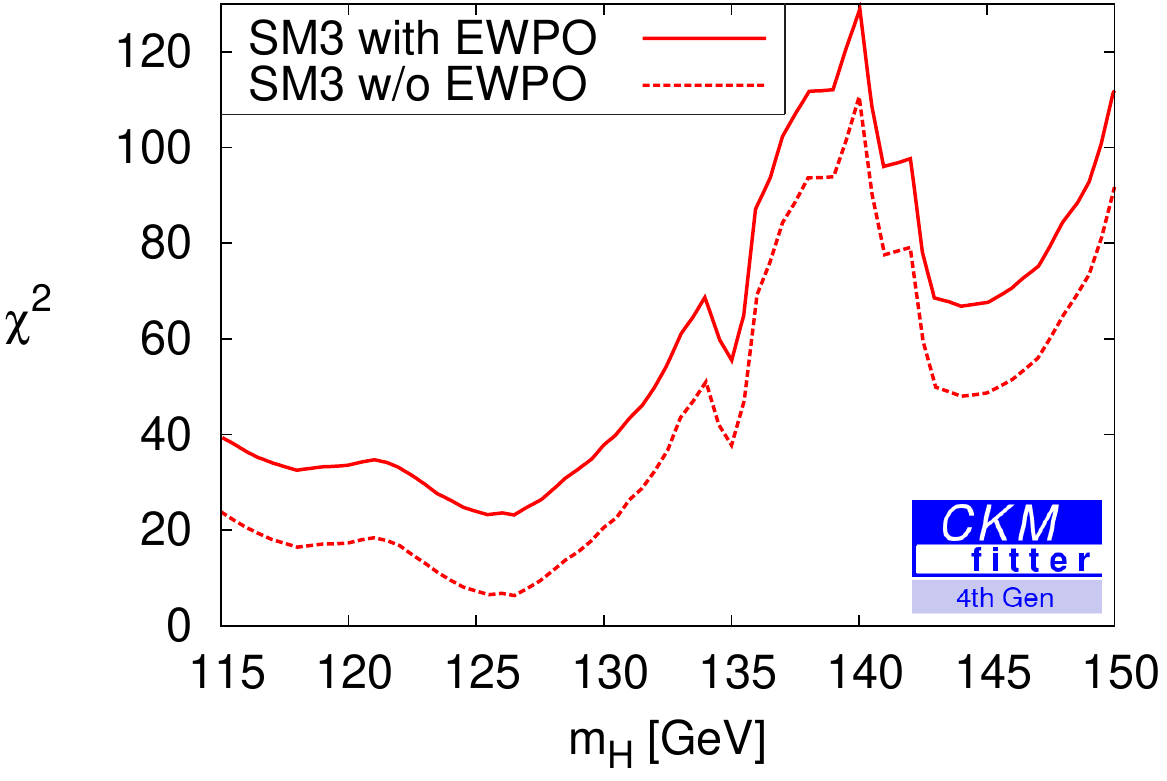}
  \caption{The minimum $\chi^2$ value of the SM3 as a function of the Higgs mass
    $m_H$. The solid line shows the results of the combined analysis of signal
    strengths and EWPOs. For the dashed line only the signal strengths were
    included in the fit.}
  \label{fig:mHSM3}
\end{figure}

\begin{figure}
  \includegraphics[width=0.45\textwidth,viewport=150 480 486 705,clip]{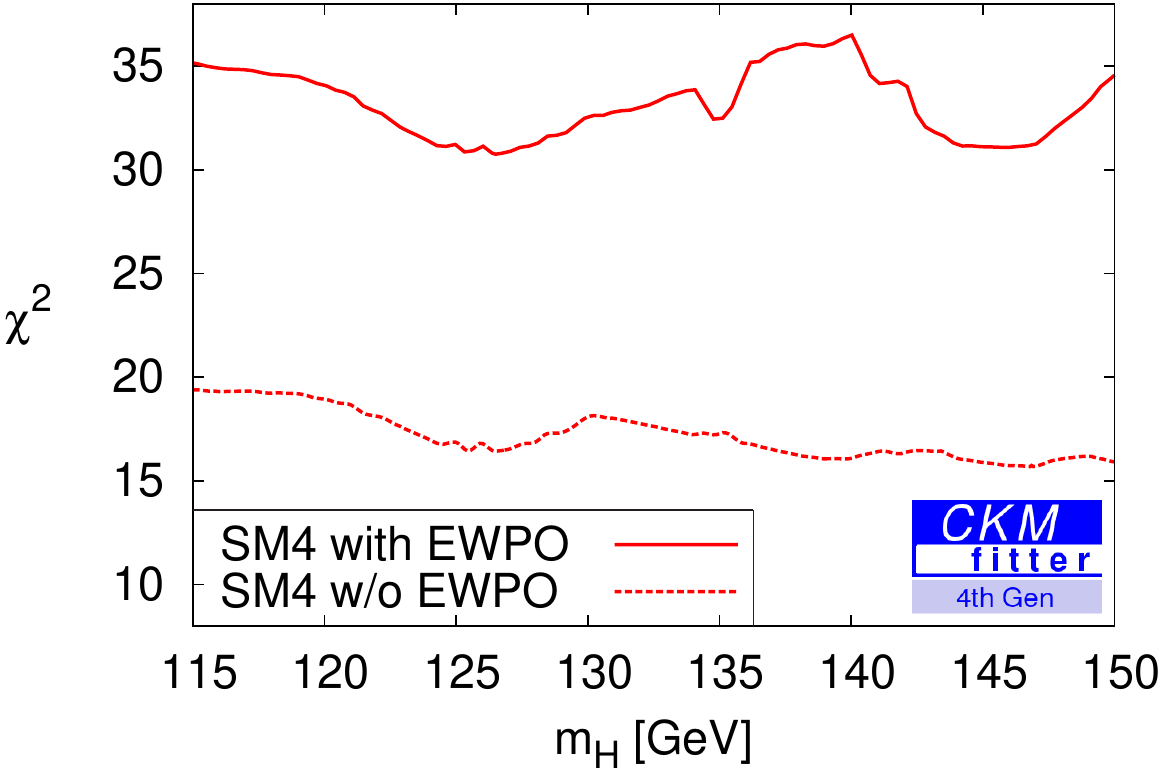}
  \caption{The minimum $\chi^2$ value of the SM4 as a function of the Higgs mass
    $m_H$. The solid line shows the results of the combined analysis of signal
    strengths and EWPOs. For the dashed line only the signal strengths were
    included in the fit.}
  \label{fig:mHSM4}
\end{figure}

Figs.~\ref{fig:mHSM3} and \ref{fig:mHSM4} show the minimum $\chi^2$ value as a
function of the Higgs mass $m_H$ in the SM3 and SM4, respectively.  The solid
lines show the results of the combined analysis of signal strengths and EWPOs
while for the dashed lines only the Higgs signal strengths (including
$H\to\tau\tau$) were used as inputs. We see that the SM3 clearly prefers a Higgs
mass near $126.5\gev$.  This is in agreement with a similar analysis presented
in \cite{Erler:2012uu}. There is another local minimum at $m_H=145\gev$, but
with a considerably larger $\chi^2$ value.  The $\chi^2$ function of the SM4 in
the combined analysis of signal strengths and EWPOs also has one minimum at
$m_H=126.5\gev$ and another one at $m_H=147\gev$. Here, the $\chi^2$ values are
almost the same, but still larger than the minimal $\chi^2$ value of the SM3,
$\chi^2 =23.3$, by about 8 units.  Note that for non-nested models or models
with bounded parameters the relation between $\chi^2$ values and p-values is no
longer given by Wilks' theorem. Thus, in the case of the SM4, the number of
degrees of freedom is an ill-defined concept and the $p$-values have to be
calculated by numerical simulation. For the simulations we used the {\myFitter}
package \cite{myFitter-page}. Further details on the statistical issues and the
{\myFitter} simulation method can be found in \cite{myFitter}.  For
$m_H=147\gev$ the signals at invariant masses near $126.5\gev$ would be
interpreted as statistical fluctuations.  Then the data would be better
described by the SM4 because it has more mechanisms for suppressing its Higgs
signals. These mechanisms were discussed in \cite{Eberhardt:2012sb}.

\begin{figure}[t]
\includegraphics[width=0.45\textwidth,bb=153 494 477 735,clip=true]{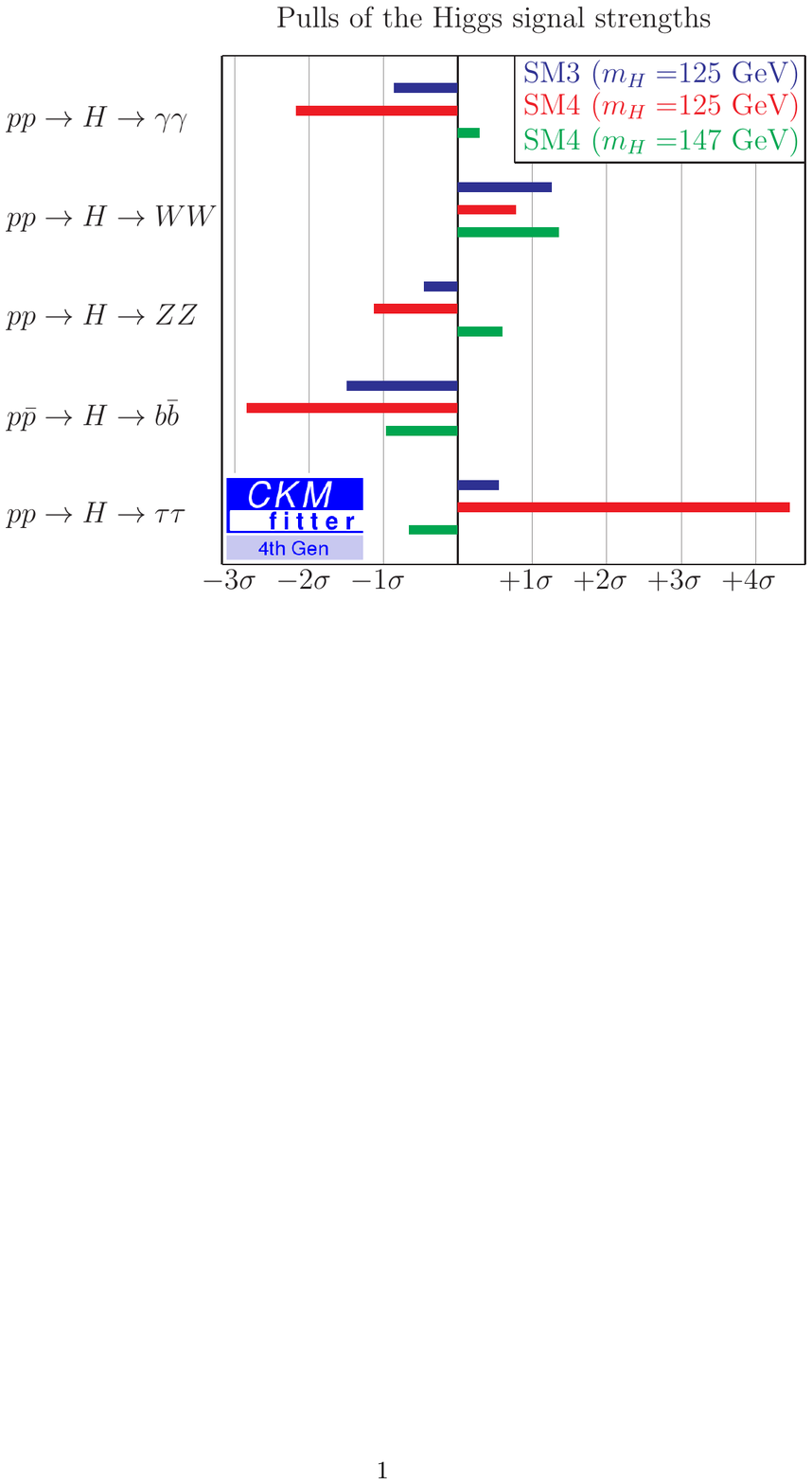}
  \caption{Pulls of the Higgs signal strengths for the SM3 with
    $m_H=126.5\gev$ and the SM4 with $m_H=126.5\gev$ and
    $m_H=147\gev$.}
  \label{fig:pulls}
\end{figure}

Fig.~\ref{fig:pulls} shows the pulls of the Higgs signal strengths for the SM3
with a Higgs mass of $126.5\gev$ and the SM4 with a Higgs mass of $126.5\gev$ or
$147\gev$. We see that in the SM4 with $m_H=126.5\gev$ the measured
$H\to\tau\tau$ signal strength deviates by more than $4\sigma$ from its
predicted value. This is due to the effect mentioned in the discussion of
Fig.~\ref{fig:mnu}. For the SM4 with $m_H=147\gev$ the measured signal strengths
for the invariant mass bin at $147\gev$ are in good agreement with their theory
predictions.  However, in that case the $\chi^2$ receives a large contribution
due to the fact that the measured values of the signal strengths in the
invariant mass bin at $126.5\gev$ deviate from their predicted values of zero.

Table~\ref{tab:pvalues} shows the $p$-values obtained from the likelihood ratio
tests for the two SM4 Higgs masses. We see that, based on the Higgs signal
strengths alone, the SM4 scenario with $m_H=126.5\gev$ is ruled out at almost
$4\sigma$ while the scenario with $m_H=147\gev$ is only excluded at $3\sigma$.
At a fixed Higgs mass of $126.5\gev$ the electroweak fit is actually better in
the SM4 than in the SM3. Thus, if the EWPOs are included in the fit, the
$p$-value increases to $2$ permille, which corresponds to $3.1\sigma$. The lower
bound $m_{t^\prime,b^\prime}\gtrsim 600\;$\gev\ is not essential for this
result, relaxing this bound to $m_{t^\prime,b^\prime}\gtrsim
400\;$\gev\ decreases the minimum-$\chi^2$ by $0.6$.  For the SM4 scenario with
$m_H=147\gev$ the $p$-value drops to $0.74$ permille ($3.4\sigma$). In any case,
the SM4 is excluded at more than $3\sigma$.
\begin{table}[t]
\centering
\renewcommand{\arraystretch}{1.3}
\begin{tabular}{l@{$\quad$}c@{$\quad$}c}
  \hline\hline
              & SM4 @ 126.5\gev & SM4 @ 147\gev \\
  \hline
  Higgs only  & $0.088\cdot 10^{-3}$ ($3.9\sigma$)
              & $2.4\cdot 10^{-3}$   ($3.0\sigma$)\\
  Higgs+EWPOs & $2.0\cdot 10^{-3}$   ($3.1\sigma$)
              & $0.74\cdot 10^{-3}$  ($3.4\sigma$)\\
  \hline\hline
\end{tabular}
\caption{$p$-values obtained from the likelihood ratio tests for fixed Higgs
  mass.  Both, the SM4 with $m_H=126.5\gev$ and $m_H=147\gev$ are compared to
  the SM3 with a fixed Higgs mass of $126.5\gev$. In the first row, only the
  Higgs signal strengths were used as inputs. The second row contains the
  results of the combined analysis of signal strengths and EWPOs. The number of
  standard deviations corresponding to each $p$-value are shown in parentheses.}
\label{tab:pvalues}
\end{table}

%%%%%%%%%%%%%%%%%%%%%%%%%%%%%%%%%%%%%%%%%%%%%%%%%%%%%%%%%%%%%%%%%%%%%%%%%%%%%%% 
\section{Conclusions}
%%%%%%%%%%%%%%%%%%%%%%%%%%%%%%%%%%%%%%%%%%%%%%%%%%%%%%%%%%%%%%%%%%%%%%%%%%%%%%% 

We presented a combined analysis of Higgs signal strengths and EWPOs in the
context of the Standard Model with three or four fermion generations. The SM3 is
in good agreement with the experimental data and the best-fit Higgs mass is
$126.5\gev$. The SM4, on the other hand, struggles to describe the Higgs signal
strengths measured at Tevatron and the LHC. The $\chi^2$ function of the SM4 has
two minima at $m_H=126.5\gev$ and $m_H=147\gev$ with essentially the same
$\chi^2$ value, which is larger than the minimal $\chi^2$ value of the SM3 by 8
units. The second minimum of the SM4 $\chi^2$ function occurs because the SM4
cannot reproduce the signal strengths measured at $126.5\gev$ very well, so that
an SM4 with a Higgs mass nowhere near the observed signals describes the data
equally well as an SM4 with $m_H=126.5\gev$. To quantitatively compare the
performance of the SM3 and SM4 at describing the data we performed likelihood
ratio tests for fixed Higgs masses $m_H$ of $126.5\gev$ in the SM3 and
$m_H=126.5\gev,\ 147\gev$ in the SM4. The $p$-values were computed with a new
numerical method \cite{myFitter} for likelihood ratio tests of non-nested
models. If EWPOs and signal strengths are included in the fit we find $p$-values
of $2.0\cdot10^{-3}$ and $0.74\cdot10^{-3}$, respectively, which means that the
SM4 is excluded at the $3\sigma$ level. While this result is obtained for Dirac
neutrinos, it will change only marginally for the case of Majorana neutrinos
with two fourth-generation mass eigenstates $\nu_4$, $\nu_5$: the fit to the
signal sthrengths will return the same invisible Higgs width, now corresponding
to the sum of the four decay rates $\Gamma(H\to \nu_{4,5}\nu_{4,5})$. A marginal
difference occurs once the EWPOs are included: choosing the $\nu_4$--$\nu_5$
mass splitting such that the eigenstate with the larger SU(2) doublet component
becomes heavier, one can slightly improve the quality of the electroweak
fit. The improvement is negligible, as indicated by the shallowness of the
minimum of the SM4 $\chi^2$ function in \fig{fig:mnu}. While the SM4 is under
severe pressure, a sequential fourth generation may still be viable in
conjunction with an extended Higgs sector
\cite{Hung:2010xh,BarShalom:2011zj,BarShalom:2012ms,He:2011ti}.

%%%%%%%%%%%%%%%%%%%%%%%%%%%%%%%%%%%%%%%%%%%%%%%%%%%%%%%%%%%%%%%%%%%%%%%%%%%%%%% 
\section*{Acknowledgements}
%%%%%%%%%%%%%%%%%%%%%%%%%%%%%%%%%%%%%%%%%%%%%%%%%%%%%%%%%%%%%%%%%%%%%%%%%%%%%%%

We thank Geoffrey Herbert and Heiko Lacker for important contributions to the
SM4 part of the CKMfitter code. We also thank J\'er\^ome Charles for CKMFitter
software support and very useful inputs regarding likelihood ratio tests for
non-nested models.

We acknowledge support by the DFG through grants NI1105/2-1, LA2541/1-1,
LE1246/9-1, and Le1246/10-1.

\bibliography{sm4higgs2}

\end{document}